# United Quark and Neutrino Mixing Matrices with Universal Pair of CP-Violating Phases

E. M. Lipmanov

40 Wallingford Road # 272, Brighton MA 02135, USA

**Abstract**

The Standard Model quark and neutrino mixing matrices are of independent empirical origin, but they do suggest unification. In this paper I obtained two united one-parameter quark and neutrino mixing matrices inferred from two *semi-empirical* deviation-from-mass-degeneracy (DMD) flavor rules —— quadratic DMD-hierarchy rule and Dirac-Majorana DMD-duality rule —— without use of the common exact-flavor-symmetry suggestions for that particular unification problem. One small empirical parameter quantitates the pattern of particle flavor physics. The main predictions are: *1)* hierarchical connections between the 2 large solar and atmospheric neutrino mixing angels, and the 2 small quark mixing angels, *2) universal sequence* of 14 equality relations to that one-empirical-parameter of the quark and neutrino mixing-matrix parameters, CP-phases and lepton mass ratios, which are free dimensionless constants in the Standard Model, *3)* complementarity connections between doubled large neutrino and small quark mixing angles, *4)* tentative solution of the CP-violation problem in framework of Standard Model mixing matrix phenomenology by suggesting a universal set of two nonzero values ~58.8$^0$ and ~31.2$^0$ for Dirac and Majorana CP-violating phases.

## 1. Introduction

The Standard Model Cabibbo-Kobayashi-Maskawa quark mixing matrix and Pontecorvo-Maki-Nakagava-Sakata neutrino one are of independent empirical origin, but they do suggest unification. One of the unsolved problems in flavor elementary particle physics is the empirical CP-violation fact. It calls for three flavor generations. The Dirac CP-violating phase in the phenomenological mixing matrix of three quark generations was first introduced in ref. [1].

In previous publications [2] and [3] I considered an empirical probably basic universal quadratic hierarchy rule for deviation-from-mass-degeneracy (DMD) physical flavor DMD-quantities that also calls for three elementary particle generations, and determines the *patterns* of most known dimensionless flavor quantities such as charged lepton (CL) and suggested quasi-degenerate (QD) neutrino mass ratios and large neutrino and small quark mixing angles and their relations. The absolute values of all these quantities are expressed through one remarkable empirical universal small physical parameter $\alpha_o \cong e^{-5} \cong 0.0067$ that is close to the free in electroweak theory fine structure constant $\alpha$ at photon pole value of momentum transfer.

After discussing the main results, I summarize them in two explicit neutrino and quark mixing matrices and show that the same quadratic flavor DMD-hierarchy rule may determine a universal equation for the Dirac and Majorana phases $\delta_n$ in the quark and lepton mixing matrices with two solutions $\delta_1$ and $\delta_2$ not based on any other empirical parameters. That origin of CP-violation is independent of the important unsolved problem of Dirac or Majorana neutrino type and suggests quantitative unification of CP-violating phases in the mixing matrixes of quarks and Dirac-or-Majorana neutrinos in addition to the unification of the quark and neutrino mixing angles.

The main result is the (approximate) quantitative low energy one-empirical-parameter *pattern* that unites all considered 'deviation-quantities' in flavor physics. The actual overall particle flavor pattern is visualized as a *small deviation* from the conceptual flavor pattern at the parameter limit $\alpha_o = 0$. Without established particle flavor theory, the considered semi-empirical deviation-from-mass-degeneracy phenomenology is suggestive and in spirit of the physics traditions.

## 2. Quadratic DMD-hierarchy flavor rule

From analyses of experimental data of particle mass ratios and mixing angles in flavor physics I suggested in [2, 3] a universal quadratic hierarchy equation for physical DMD-quantities in the form

$$[DMD(2)]^2 \cong 2[DMD(1)] \qquad (1)$$

where DMD(n), n=1,2, denote deviations from unity of the relevant particle flavor dimensionless quantities: 1) particle mass ratios squared and 2) particle mixing parameters. The hierarchy rule (1) answers the specific quantitative neutrino-quark problem of two empirically large hierarchical solar and atmospheric mixing parameters $Sin^2 2\theta_{12}$ and $Sin^2 2\theta_{23}$ and its relation to two small quark mixing parameters. The particular hierarchy equations for neutrino and quark mixing parameters are given [3] by

$$(Sin^2\ 2\theta_{12} -1)^2\ \cong\ 2\left|Sin^2\ 2\theta_{23} -1\right|,$$

$$(Cos^2\ 2\theta_c - 1)^2\ \cong\ 2\left|Cos^2\ 2\theta' - 1\right|, \tag{2}$$

or

$$Cos^2\ 2\theta_{12}\ \cong\ \sqrt{2}\ Cos\ 2\theta_{23},$$

$$Sin^2\ 2\theta_c\ \cong\ \sqrt{2}\ Sin\ 2\theta', \tag{2'}$$

where $\theta_c$ is the Cabibbo angle and $\theta'$ is the next to the largest quark mixing angle. The DMD(n)-quantities are interpreted in (2) as *deviations* from maximal or minimal mixing for neutrinos or quarks respectively.

The equations (2), or (2'), as illustrations of the significant contents of the quadratic hierarchy rule (1), are independent of empirical parameters; they are directly measurable pertinent relations between generic *doubled* mixing angles of the neutrino and quark mixing matrixes – the atmospheric neutrino mixing parameter is determined by the more accurate value of the solar mixing parameter, and the $\theta'$-angle is determined by the value of the Cabibbo angle [3]. Results are in good agreement with neutrino oscillation data [6] and the Wolfenstein [4] parameterization of the quark CKM-mixing matrix.

There are four interesting solutions of the hierarchy Eq.(1) with dual, large and small, extended DMD-quantities. The solution of Eq.(1) for large DMD-quantities is for CL-mass ratios[1]

$$(m_\tau/m_\mu)^2\ \cong\ 2/\alpha_o,\ \ (m_\mu/m_e)^2 \cong\ 2/\alpha_o{}^2,\ \ \alpha_o = e^{-5}. \tag{3}$$

Three solutions of Eq.(1) are obtained for small DMD-quantities. 1) QD-neutrino mass ratios

$$(m_3{}^2/m_2{}^2)\ \cong\ \exp(2r),\ \ (m_2{}^2/m_1{}^2) \cong \exp(2r^2), \tag{4}$$

---

[1] The quark mass ratios are not considered explicitly here.

where $(m_1 < m_2 < m_3)$ are the neutrino masses and $r$ is the neutrino oscillation solar-atmospheric hierarchy parameter $r = (m_2^2 - m_1^2)/(m_3^2 - m_2^2) \cong 5\alpha_o$.

2) Small deviations of large neutrino mixing parameters from maximal mixing as solution of Eq,(2) in close agreement with oscillation data [3],

$$\text{Sin}^2\ 2\theta_{12} \cong (1 - 2\sqrt{\alpha_o}) \cong 0.836,\ \ 2\theta_{12} \cong 66.1°,$$

$$\text{Sin}^2\ 2\theta_{23} \cong (1 - 2\alpha_o) \cong 0.987,\ \ 2\theta_{23} \cong 83.5°, \qquad (5)$$

in good agreement with best-fit oscillation data [6].

3) Almost equally small deviations of quark mixing parameters from minimal mixing [3],

$$\text{Cos}^2\ 2\theta_c \cong (1 - 2\sqrt{\alpha_o}) \cong 0.836,\ \ 2\theta_c \cong 23.9°,$$

$$\text{Cos}^2\ 2\theta' \cong (1 - 2\alpha_o) \cong 0.987,\ \ 2\theta' \cong 6.6°. \qquad (6)$$

I should emphasize here that the involved parameter $\alpha_o$ does not enter in the primary quadratic hierarchy equation (1) and appears, completely legitimate in physics as experimental science, by empirical universal hints from experimental data such as: 1) CL mass ratios (3), 2) relation between mass ratios of QD-neutrinos and small value of the solar-atmospheric hierarchy parameter (4), 3) two large neutrino mixing angles (5), and 4) two small quark mixing angles (6). An interesting empirical suggestion is the *dual* relations between large CL DMD-quantities and small QD-neutrino DMD-quantities, and between the two pairs of large neutrino and small quark mixing angles. If the neutrinos are Majorana particles, that dual relation may be more generally described by the term 'Dirac-Majorana DMD-duality'.

For illustration of the Dirac-Majorana DMD-duality between large and small DMD-solutions in (3)-(6), consider the virtual limit $\alpha_o \to 0$: the divergence of CL masses gets infinitely large and the CL mixing (if present) disappears, the neutrinos get exactly mass-degenerate and neutrino mixing is getting maximal, the divergences of quark mass spectra are getting infinitely large[2] and the quark mixing disappears.

This line of reasoning explains why the elements $S^{nu}_{13}$ and $S^{q}_{13}$ in the neutrino and quark mass matrices respectively, supposed to be equal to each other (self-dual), should be very small: by (3) and (4), in the limit $\alpha_o = 0$ the neutrinos should be exactly mass-degenerate versus infinitely divergent CL masses and therefore CP-symmetry is conserved; with preserved continuity, it means that in the actual case of finite but small parameter $\alpha_o$ CP-violation should be small, the matrix elements $S^{nu}_{13}$ and $S^{q}_{13}$ should be small but finite independently of the CP-violating exponential factors $e^{i\delta}$, and their values may be also related to $\alpha_o$

$$S^{q}_{13} \cong S^{nu}_{13} \cong \alpha_o/2 \cong 0.0034, \qquad (7)$$

as hinted by the quark experimental data [6].

With the arguments above, we get the explicit quantitative set of the neutrino mixing matrix $V_\ell$ with Dirac CP-violating phase $\delta_\ell$ in the standard parameterization [6]

$$V_\ell \cong \begin{pmatrix} c_{12} & s_{12} & s_{13}e^{i\delta} \\ -s_{12}c_{23} & c_{12}c_{23} & s_{23} \\ s_{23}\,s_{12} & -s_{23}c_{12} & c_{23} \end{pmatrix}_\ell \cong \begin{pmatrix} 0.83 & 0.55 & 0.0034\exp(i\delta_\ell) \\ -0.41 & 0.62 & 0.67 \\ 0.37 & -0.55 & 0.75 \end{pmatrix}, \qquad (8)$$

---

[2] The particular suggestion of quark-CL mass-spectrum analogy is maintained here, it is supported by the large empirically know divergence of quark mass spectra.

where

$$C_{12} = \sqrt{[\tfrac{1}{2} + \sqrt{(\sqrt{\alpha_o}/2)}]}, \quad S_{12} = \sqrt{[\tfrac{1}{2} - \sqrt{(\sqrt{\alpha_o}/2)}]},$$

$$C_{23} = \sqrt{[\tfrac{1}{2} + \sqrt{(\alpha_o/2)}]}, \quad S_{23} = \sqrt{[\tfrac{1}{2} - \sqrt{(\alpha_o/2)}]},$$

$$(S_{13}\,e^{i\delta})_\ell = (\alpha_o/2)\exp(i\delta_\ell). \tag{9}$$

Compare the neutrino mixing matrix (8) with the widely discussed tribimaximal Harrison-Perkins-Scott (HPS) [5] mixing matrix with maximal atmospheric neutrino mixing

$$\cdot \begin{pmatrix} \sqrt{2}/\sqrt{3} & 1/\sqrt{3} & 0 \\ -1/\sqrt{6} & 1/\sqrt{3} & 1/\sqrt{2} \\ 1/\sqrt{6} & -1/\sqrt{3} & 1/\sqrt{2} \end{pmatrix}$$

The deviation of the atmospheric neutrino oscillation parameter $S_{23}$ in (8) from the maximal-mixing-value in HPS is small, but not zero ~6%. The deviation of the solar neutrino oscillation parameter $S_{12}$ in (8) from the HPS value is ~5%, while its deviation from maximal mixing value is ~28%.

The quark mixing matrix $V_q$, by comparison of (5) and (6), follows from the lepton one (8) by the interchange $\mathrm{Sin}\,2\theta_{12} \to \mathrm{Cos}\,2\theta_c$, $\mathrm{Sin}\,2\theta_{23} \to \mathrm{Cos}\,2\theta'$

$$V_q \cong \begin{pmatrix} C_{12} & S_{12} & S_{13}e^{i\delta} \\ -S_{12}C_{23} & C_{12}C_{23} & S_{23} \\ S_{23}\,S_{12} & -S_{23}C_{12} & C_{23} \end{pmatrix}_q \cong \begin{pmatrix} 0.98 & 0.21 & 0.0034\exp(i\delta_q) \\ -0.21 & 0.98 & 0.058 \\ 0.01 & -0.06 & 0.998 \end{pmatrix},$$

$$\tag{10}$$

with notations

$$(C_{12})_q = \sqrt{[\tfrac{1}{2} + \tfrac{1}{2}\sqrt{(1-2\sqrt{\alpha_o})}]}, \quad (S_{12})_q = \sqrt{[\tfrac{1}{2} - \tfrac{1}{2}\sqrt{1-2\sqrt{\alpha_o}}]},$$

$$(C_{23})_q = \sqrt{[\tfrac{1}{2} + \tfrac{1}{2}\sqrt{(1-2\alpha_o)}]}, \quad (S_{23})_q = \sin\theta' \cong \sqrt{[\tfrac{1}{2} - \tfrac{1}{2}\sqrt{(1-2\alpha_o)}]},$$

$$(S_{13}\,e^{i\delta})_q = (\alpha_o/2)\exp(i\delta_q). \tag{11}$$

So, we obtained two *one-parameter* elementary particle mixing matrices (8) and (10) for the leptons and quarks respectively.

I should notice, that all the relations between the neutrino and quark mixing matrix elements and the parameter $\alpha_o$ in the lists (9) and (11) follow from only *one* simple empirical connection $\cos^2 2\theta_{23} \cong 2\alpha_o$, or $\sin^2 2\theta_c \cong 2\sqrt{\alpha_o}$ etc, and the quadratic hierarchy and DMD-duality flavor rules. So, the reasonable agreement of all other lepton and quark mixing matrix elements (8) and (10) with data is a supporting experimental test of these two flavor rules. The conditions of unitarity of the mixing matrices (8) and (10) are fulfilled with accuracy better than 1%.

The cause of small CP-violation in the elementary particle mass matrix phenomenology is the small value of the element $s_{13}$ independent of the phase values. The lepton and quark Dirac CP-violating phases $\delta_l$ and $\delta_q$ are phase-convention independent, but still remain undetermined and may be different.

If the phases $\delta_l$ and $\delta_q$ are equal zero, the CP-symmetry is conserved. Therefore, the problem of empirically discovered but not explained CP-violation in the framework of the Standard Model mixing matrix phenomenology is reduced to the problem of discovering the source of the nonzero numerical values of CP-violating phases. A solution to this problem from the quadratic hierarchy generic DMD-flavor rule (1) is considered below.

### 3. Dirac-Majorana DMD-duality flavor rule

The main physical quantities in the present flavor phenomenology are the deviation-from-mass-degeneracy DMD(n)-quantities, n=1,2, for generic flavor pairs. In case of lepton mass ratios they are:

1) large DMD-quantities of CL mass ratios squared

$$[(m_\tau/m_\mu)^2 - 1] \cong 2/\alpha_o, \quad [(m_\mu/m_e)^2 - 1] \cong 2/\alpha_o^2, \qquad (3')$$

2) small DMD-quantities of QD-neutrino mass ratios squared

$$[(m_3/m_2)^2 - 1] \cong 2r, \quad [(m_2/m_1)^2 - 1] \cong 2r^2, \quad r \cong 5\alpha_o. \qquad (4')$$

The magnitudes of the pairs of dual deviations from maximal and minimal mixing in (3') and (4') are 2a and $2a^2$ where $a = 1/\alpha_o >> 1$ or $a = r << 1$ for CL and neutrinos respectively.

The DMD-pairs (3') and (4') approximately describe the deviation of the CL and neutrino mass patterns from their limiting form at $\alpha_o = 0$ to the actual observable finite lepton mass pattern at finite small value of the parameter $\alpha_o \cong 0.0067$ – that small $\alpha_o$-shift produces infinitely large shift of CL masses and only a small shift of neutrino masses.

Each of the generic flavor pairs (3') and (4') of lepton DMD-quantities obey the quadratic hierarchy flavor rule (1). At the same time these large and small lepton DMD-pairs obey the second flavor rule of Dirac-Majorana DMD-duality (neutrinos are supposed Majorana particles) -- the real CL mass-ratios are large as remnants of infinitely divergent ones at $\alpha_o = 0$, while neutrino mass-ratios are ~1 as small deviation from exact mass-degeneracy at that limit.

Another two 'large' and small generic pairs of *extended* flavor DMD-quantities introduced above are

3) 'large' neutrino-mixing DMD-quantities

$$[1 - Cos^2 2\theta_{12}] \cong (1 - 2\sqrt{\alpha_o}) , \quad [1 - Cos^2 2\theta_{23}] \cong (1 - 2\alpha_o) , \quad (5')$$

4) small quark-mixing DMD-quantities

$$[1 - Cos^2 2\theta_c] \cong (2\sqrt{\alpha_o}) , \quad [1 - Cos^2 2\theta'] \cong (2\alpha_o) . \quad (6')$$

If considered as deviations from maximal mixing for neutrinos and minimal mixing for quarks, see (5) and (6), each of the generic DMD-pairs (5') and (6') obey the quadratic hierarchy rule (1) (in the horizontal lines). At the same time those quark and neutrino DMD-pairs obey the Dirac-Majorana DMD-duality flavor rule (in the vertical lines), but in a special form: 'large' means here ~1. In the limit $\alpha_o = 0$ the neutrino mixing is exactly maximal $Sin^2 2\theta_{ij} = 1$ while the quark mixing parameters disappear. At the finite actual small value $\alpha_o \cong 0.0067$ the observable neutrino mixing is a small deviation from maximal value 1 versus quark mixing a small deviation from minimal (zero) value. The neutrino-quark duality connections (5) and (6) predict quark-neutrino complementarity mixing relations [8]:

$$Cos^2 2\theta_{12} \cong Sin^2 2\theta_c, \quad Cos^2 2\theta_{23} \cong Sin^2 2\theta',$$

$$2\theta_{12} \cong (\pi/2 - 2\theta_c), \quad 2\theta_{23} \cong (\pi/2 - 2\theta'). \quad ((5) + (6))$$

To summarize, Dirac-Majorana DMD-duality unites large neutrino and small quark parameters in the low energy Standard Model mixing matrices. It unites the DMD-quantities for mass ratios of QD-neutrinos and charged leptons. And that duality is the reason why the new dimensionless small parameter $\alpha_o$ is needed in flavor physics, see Sec. 6.

I should noticed, that neutrino and quark mixing matrix unification at low energies was obtained earlier by Mohapatra et al [7] from the suggestion of equal CKM-mixings for both quarks and leptons at seesaw scale with radiative corrections generating the observed large mixing angles for degenerate neutrinos, but keeping the small CKM ones for quarks due to their mass hierarchy.

## 4. Quadratic hierarchy flavor rule as origin of Dirac CP-violating phase values

Dirac CP-violating phases enter the mixing matrixes of quarks $V_{CKM}$ and neutrinos $V_{nu}$ in the form of the exponential factors $e^{i\delta}$. As of to date, there is no known connection between the two Dirac phases for quarks $\delta_q$ and neutrinos $\delta_\ell$. Such a connection is suggested by the universal quadratic hierarchy equation in flavor physics and considered below.

The real and imaginary parts, $\cos\delta$ and $\sin\delta$, of the exponential factor $e^{i\delta}$ are a pair of two generic flavor quantities related to three particle generations and should obey the hierarchy equation (1), (2), for generic flavor pairs. Since in this peculiar case both of the generic quantities depend on the same unknown $\delta$-parameter, the hierarchy Eq.(1) should determine these parameters uniquely.

The physical meaning of Eq.(1) for the $\delta$-angles is that it defines a quadratic hierarchy relation between the deviations from maximal and minimal CP-violating phase values $(1-\sin^2\delta)$ and $(1-\cos^2\delta)$ respectively. Since its solution is symmetric under exchanges of $\sin^2\delta$ and $\cos^2\delta$, the two solutions must be complementary

$$\delta_1 + \delta_2 = \pi/2. \tag{12}$$

Consider the two sole possibilities:

1). $Sin^2\delta_1 > Cos^2\delta_1$. In that case, DMD(2)=$(1 - Cos^2\delta_1)$, DMD(1)=$(1 - Sin^2\delta_1)$, and from (1) follows a definite equation for the phase $\delta_1$,

$$(1 - Cos^2\delta_1)^2 \cong 2(1 - Sin^2\delta_1), \qquad (13)$$

or

$$Sin^2\delta_1 \cong \sqrt{2} Cos \delta_1. \qquad (13')$$

Its only solution is

$$Cos \delta_1 \cong \sqrt{(2 - \sqrt{3})} \cong 0.518, \ \delta_1 \cong 58.8^o. \qquad (14)$$

2). $Sin^2\delta_2 < Cos^2\delta_2$, DMD(2)$\leftrightarrow$DMD(1), and the equation for $\delta_2$ is given by

$$Cos^2\delta_2 \cong \sqrt{2} Sin \delta_2, \qquad (15)$$

with the solution

$$Sin \delta_2 \cong 0.518, \ \delta_2 \cong 31.2^o. \qquad (16)$$

Those two values for the Dirac phases $\delta$ are the only solutions of the hierarchy equation (1) independent of any adjusting empirical parameters.

The first solution $\delta_1 \cong 58.8^o$ is well within the experimental data ranges of the CP-violating phase of the quark CKM-mixing matrix [6], $\delta_{CKM} = 59^o \pm 13^o$. So, the obtained value $\delta_1$ should be the prediction for the quark CP-violating phase

$$\delta_q \cong 58.8^o. \qquad (17)$$

Since the two values for the Dirac phase are uniquely connected by the universal hierarchy rule, it is natural to identify the second solution $\delta_2 \cong 31.2^o$ with the unknown value of the CP-violating Dirac phase $\delta_l$ of the neutrino mixing matrix

$$\delta_l \cong 31.2^o. \qquad (18)$$

The value (18) is the predicted Dirac CP-violating phase in the neutrino mixing matrix.

To summarize, the quadratic hierarchy equation as a rule in flavor physics may be the origin of the Dirac phases in quark and lepton mixing matrixes and predicts: 1) necessary CP-violation in the quark mixing matrix – not vanishing Dirac phase $\delta_q \neq 0$, 2) definite and not maximal or minimal values (17) and (18) for the Dirac CP-violating phases in the united quark and lepton mixing matrixes, 3) the pair of quark and lepton Dirac phase values is fully determined by the hierarchy rule (1) in contrast to the pairs of mixing angle values related also to the parameter $\alpha_o$, 4) quark-lepton complementarity[3] is a common feature of the (1-2) and (2-3) mixing angles *and Dirac CP-violating phases*.

## 5. Universal pair of CP-violating phases in quark and Majorana neutrino mixing matrices

The unique simple phenomenological explanation of the discovered by Lee, Yang and Wu P-symmetry violation is introduced by Feinman, Gell-Mann and Marshak, Sudarshan in the form of universal empirical (V – A)x(V - A) lepton and quark weak interaction rule. An analogous universal explanation of the discovered in 1964 by Christenson, Cronin, Fitch and Turlay [8] CP-violation in the decays of K-mesons is not known at present.

Since the values of most low energy particle dimensionless flavor quantities are described by the hierarchy rule (1) as considered above, and in accordance with good traditions in basic frontier physics serving experimental search, I suggest a universal CP-violation mechanism: the obtained above one unique pair of nonzero phase values $\delta_1 \cong 58.8^{\circ}$ and $\delta_2 \cong 31.2^{\circ}$ may be the universal simple source of CP-violation at least in the Standard Model mixing matrix phenomenology of elementary particle interactions. Without data indications to the contrary, such a scheme with minimal CP-violating parameters is a possible unification of exponential phase factors deduced from the hierarchy rule (1). The condition that CP-violating phases come in pairs is fulfilled in the interesting case of Majorana neutrinos in the framework of three particle generations. So, the pair of two Majorana phases[4] $\alpha^{M}_{21}$ and $\alpha^{M}_{31}$, by the same motivation as the pair of quark and neutrino Dirac phases $\delta_q$ and $\delta_{\ell}$, should be chosen from that universal pair of values ~58.8$^{\circ}$ and ~31.2$^{\circ}$. Note, that the significance of Majorana phases in neutrino oscillation experiments was first demonstrated by Schechter and Valle [10].

---

[3] Complementarity relations between large lepton and small quark mixing angles were introduced in [9].

As a result, there are two choices for two different Majorana neutrino phases $\alpha^{M}_{21} \cong 58.8^{\circ}$ and $\alpha^{M}_{31} \cong 31.2^{\circ}$, or $\alpha^{M}_{21} \cong 31.2^{\circ}$ and $\alpha^{M}_{31} \cong 58.8^{\circ}$, to be decided from analyses of coming experimental data.

## 6. Flavor physics with one empirical parameter $\alpha_{\circ}$

In the previous Sections it is shown, that all quantities which enter in the low energy lepton and quark mixing matrices, except the one universal pair of CP-violating phases, can be with fair approximation expressed through only one empirical parameter $\alpha_o$.

---

[4] For notations see [10, 11].

Here I show that the remaining CP-violating phase values can be incorporated in the one-$\alpha_o$-parameter flavor pattern to within the same fair approximation.

From the obtained phase values (14), (16) and (12), relations for the CP-violating phases $\delta_1$ and $\delta_2$ follow

$$(\delta_1 - \pi/4) = (\pi/4 - \delta_2) \cong \theta_c. \qquad (19)$$

Then, from (19) and (5), (6) we get

$$\text{Cos}^2 2\delta_1 = \text{Cos}^2 2\delta_2 \cong \text{Sin}^2 2\theta_c \cong \text{Cos}^2 2\theta_{12} = (2\sqrt{\alpha_o}), \quad \delta_2 \cong \theta_{12}. \quad (20)$$

By these relations, the deviation of the phase $\delta_2$ from maximum mixing angle value $\theta_m = \pi/4$ is related to the value of the Cabibbo mixing angle $\theta_c \cong 13^{\,o}$; the values of the two CP-violating angles are shifted from $\theta_m = 45^\circ$ by the Cabibbo angle $\theta_c$:

$$\delta_1 \cong (\pi/4 + \theta_c), \quad \delta_2 \cong (\pi/4 - \theta_c). \qquad (19')$$

The interesting physical meaning of the statements (19) and (20) if added to the discussed results above is that all dimensionless flavor physical quantities in the neutrino, CL and quark *mixing matrices* are expressed with fair approximations (reserved for probable specifications) through the one empirical universal parameter $\alpha_o$ revealing an interesting simple *physical pattern*. All 10 values of mixing angles and Dirac and Majorana CP-violating phases in the standard parameterization of neutrino and quark Standard Model mixing matrices $V_\ell$ and $V_q$ are approximately presented by two equality sequences:

$$2\sqrt{\alpha_o} \cong \text{Cos}^2 2\theta_{12} \cong \text{Sin}^2 2\theta_c \approx \text{Cos}^2 2\delta_\ell \cong \text{Cos}^2 2\delta_q$$
$$\cong \text{Cos}^2 2\alpha^{\textbf{M}}{}_{21} \cong \text{Cos}^2 2\alpha^{\textbf{M}}{}_{31} \cong 2\sqrt{2}\ (m_\mu/m_\tau), \qquad (21)$$

$$2\alpha_o \cong \mathrm{Cos}^2 2\theta_{23} \cong \mathrm{Sin}^2 2\theta' \cong 2\,\mathrm{Sin}\, 2\theta^q{}_{13}$$

$$\cong 2\,\mathrm{Sin}\, 2\theta^l{}_{13} \cong 2\sqrt{2}\ (m_e/m_\mu)\,. \tag{22}$$

These two sequences show that (i) all elements of the neutrino and quark mixing matrices and CL mass ratios are united by one universal empirical parameter $\alpha_o$ -- a total of 12 *free* in the Standard Model dimensionless parameters are quantitatively related by the sequences (21) and (22), (ii) the two sequences of flavor quantities are connected by the quadratic DMD-hierarchy equation (1), (iii) all 10 relations between the mixing angles and CP-violating phases of neutrino and quark mixing matrices and parameter $\alpha_o$ in (21) and (22) follow from the two quadratic DMD-hierarchy and Dirac-Majorana DMD-duality flavor rules (2) and only one empirical relation, e.g. $\mathrm{Cos}^2 2\theta_{23} \cong 2\alpha_o$.

Consider that *flavor pattern* at the virtual limit $\alpha_o \to 0$: *1)* The divergence of CL masses gets infinitely large, *2)* the neutrinos are getting exactly mass-degenerate, *3)* two large neutrino mixing angles are getting maximal $\theta_{12} = \theta_{23} = \theta_m \equiv 45^o$, *4)* the small neutrino mixing angle gets negligibly small $\theta_{13} = 0$, *5)* all four Dirac and Majorana lepton CP-violating phases get equal $\alpha^M{}_{21} = \alpha^M{}_{31} = \delta_l = \delta_q = \pi/4$, *6)* the divergences of quark mass spectra, like the CL ones, are getting infinitely large with all three quark mixing angles getting negligibly small and the CP-violating phase approaching the value $\delta_q = \pi/4$, *7)* all remaining not zero *finite* mixing angles and CP-violating Dirac and Majorana phases in the neutrino and quark mixing matrices are getting equal to one limiting value $\theta_m = 45^o$ in fair agreement with neutrino-quark complementarity relations [9] and relations (19) for the CP-violating phases:

$$\theta_{12} \cong (\pi/4 - \theta_c), \ \theta_{32} \cong (\pi/4 - \theta'),$$

$$\delta_1 \cong (\pi/4 + \theta_c), \ \delta_2 \cong (\pi/4 - \theta_c), \tag{23}$$

with $\theta_c$ and $\theta'$ related to $\alpha_o$ by Eq.(6). So, the angle $\theta_m = 45^o$ is special in the phenomenology of lepton mixing matrices with CP-violating phases – the physically observable atmospheric and solar neutrino mixing angles and the CP-violating phases are grouped around that value with the *deviations from it expressed by the small quark mixing angles* $\theta_c$ *and* $\theta'$ *in (23).*

So, the simple physical meaning of the puzzling empirical fact of two large neutrino mixing angles versus two small quark ones is: large neutrino mixing angles are related to exact neutrino mass-degeneracy in the virtual limit $\alpha_o \to 0$, whereas small quark mixing angles are related to infinitely divergent quark masses in that limit. The overall real physical low energy lepton and quark flavor pattern is a 'small' deviation from the considered virtual flavor pattern generated by the small deviation from zero of the finite value of the empirical parameter $\alpha_o \cong 0.0067$. It is an approximate but physically attractive one-parameter quark and lepton flavor pattern which unites 14 free in the Standard Model dimensionless flavor parameters – 6 mixing angles, 4 CP-violating phases, 2 CL mass ratios, and 2 QD-neutrino mass ratios. For comparison it should be mentioned that the limiting case of exact mass-degeneracy of Majorana neutrinos was studied in [12] and shown that only two mixing angles and one Majorana phase survive in that limit.

Why are deviations from the virtual simple flavor pattern ('conceptual' flavor pattern without small parameters) really needed? The answer is the Dirac-Majorana DMD-duality condition. Indeed, in case of exactly mass-degenerate neutrinos this duality requires that the mass spectra of CL and quarks must be infinitely divergent, but that is impossible in physics on general grounds[5]. On the other hand, with *quasi-degenerate* Majorana neutrinos the Dirac particle mass spectra should have large but finite divergences in conformity with physical facts. It is clear that such deviation from the simple conceptual flavor pattern requires minimum one new dimensionless small physical parameter to measure the scale of neutrino mass-degeneracy violation. That is why the universal parameter $\alpha_o$ is really needed; it established the bare finite physical flavor DMD-quantities. That such goal may be achieved with only one parameter is a remarkable suggestion from the low energy experimental particle flavor data.

To summarize, the parameter $\alpha_o$ in present flavor phenomenology is a universal quantitative measure of the small deviations of real CP-violating particle flavor *pattern* from the conceptual very simple CP-conserving one at $\alpha_o = 0$.

### 7. Conclusion

In accordance with the above results, the quadratic hierarchy rule (1) should be a significant regularity in the phenomenology of low energy elementary particle CP-violating flavor physics. The most distinct predictions from that flavor rule, supplemented by quark-neutrino (or likely Dirac-Majorana) DMD-duality rule, are:

1. *Completely united neutrino and quark one-empirical-parameter approximate mixing matrices (8) and (10)* in agreement with experimental data. Conformable hierarchical connections between the two large neutrino mixing angles ($Cos^2 2\theta_{12} \cong \sqrt{2} Cos 2\theta_{23}$) and the two small quark ones ($Sin^2 2\theta_c \cong \sqrt{2} Sin 2\theta'$), and DMD-duality quark-neutrino mixing angle connections ($Cos 2\theta_{12} \cong Sin 2\theta_c$) and ($Cos 2\theta_{23} \cong Sin 2\theta'$) are particularly interesting results.

2. All physically meaningful 'deviation-quantities' in flavor physics —— deviations of the two large atmospheric and solar neutrino mixing angles from maximal mixing (5), deviations of the two small quark mixing angles from minimal mixing (6), deviations of (1-3) neutrino and quark mixing angles from zero (7), deviations of the Dirac and Majorana CP-violating phases from the special $45^o$-value (23) and deviations of CL and neutrino mass ratios from unity —— are expressed through one small universal empirical parameter $\alpha_o$ in reasonable agreement with data.

3. The actual overall elementary particle low energy flavor pattern is represented as a 'very small' deviation $\sim \Delta\alpha_o \cong 0.0067 << 1$ from the simple conceptual flavor pattern at the limiting value of parameter $\alpha_o = 0$.

4. The angles $\theta_c$ and $\theta'$ are directly observable estimates of the deviations of large neutrino mixing angles and Dirac and Majorana CP-violating phases from the special value ($45^o$) and deviations of small quark mixing angles from minimal value ($0^o$), see (23).

5. One source of CP-violation in quark and Majorana neutrino Standard Model mixing matrix phenomenology in the form of a *universal pair* of nonzero Dirac and Majorana CP-violating phases to be chosen from the set $\delta_1 \cong 58.8°$ and $\delta_2 \cong 31.2°$. That universal pair is fitting well in the system of dimensionless flavor quantities, see the two sequences (21) and (22) connected by hierarchy rule (1).

6. The addressed CP-noninvariant lepton and quark *flavor pattern* with a *special status of neutrinos* is suggestive and testable by coming experimental data, especially data on precise neutrino oscillation parameters, double neutrinoless beta-decay and Dirac and Majorana CP-violating phases.

---

[5] Especially with the condition that the mass of one particle, e.g. electron, should be finite.

I would like to thank very much J. W. F. Valle, R. N. Mohapatra and M. K. Parida for the interest and information.